\documentclass[preprint,12p,authoryear]{elsarticle}

\usepackage{graphicx}
\usepackage{amsmath}
\usepackage{comment}
\usepackage{appendix}
\usepackage{xcolor}
\usepackage{tabularx}
\usepackage[version=4]{mhchem}
\biboptions{sort&compress}
\usepackage{amssymb}
\usepackage[colorlinks=true]{hyperref}

\begin{document}

\begin{frontmatter}

\title{Turbulence impacts synthetic methanol production in a square duct flow: a direct numerical simulation (DNS) study}

\author[1]{Marko Korhonen\corref{cor1}}
\ead{marko.korhonen@aalto.fi}
\author[2]{Daulet Izbassarov}
\author[1]{Judit Nyári}
\author[1]{Annukka Santasalo-Aarnio}
\author[1]{Ville Vuorinen}

\cortext[cor1]{Corresponding author}

\affiliation[1]{organization={Department of Mechanical Engineering, Aalto University},
addressline={Otakaari 4},
postcode={P.O. Box 14100 FI-00076 AALTO},
city={Espoo},
country={Finland}}
\affiliation[2]{organization={Finnish Meteorological Institute},
addressline={Erik Palménin aukio 1},
postcode={P.O. Box 503 FI-00101 Helsinki Finland},
city={Helsinki},
country={Finland}}


\begin{abstract}
Improving the synthetic methanol production efficiency is essential for adopting carbon-neutral methanol technologies. 
In the present computational fluid dynamics (CFD) study, we investigate synthetic methanol (CH$_3$OH) production in a milli-duct with a rectangular cross-section. The duct involves catalytic wall reactions with CO$_2$ and H$_2$ as reactants. Direct numerical simulations are carried out at bulk Reynolds numbers $Re_\mathrm{b} =$ 100, 500, 1100 and 2200 spanning laminar and turbulent flow regimes. We investigate a high catalyst loading, corresponding to gas hourly space velocity values of GHSV = 0.086, 0.44, 0.96 and 1.93 m$^3$/(kg$_\mathrm{cat}$ h). The results consist of the following features. 1) Turbulent mixing may significantly enhance methanol production under transport-limited chemistry, increasing the methanol mass flow rate at the outlet up to 56.4 \% in the duct. 2) The reaction rate reaches local minima/maxima in the turbulent ejection/sweep structures observed in the proximity of the catalytic walls, respectively. 3) The single-pass conversion efficiency of \ce{CO2} and the methanol yield across the duct decrease monotonically as the Reynolds number increases (lower residence time), while in contrast, the methanol mass flow rate at the outlet increases as $Re_\mathrm{b}$ increases (higher reactant inflow). These results suggest that turbulent transport could be leveraged to improve methanol yield in straight millichannels, albeit at the cost of increased reactant recycling.
\end{abstract}


\end{frontmatter}

\section{Introduction}
As the adverse impacts of climate change continue to become more apparent worldwide, there is mounting pressure to reduce carbon emissions originating from fossil fuel use~\citep{hoegh2019human}. Therefore, the transition to renewable energy resources is considered essential~\citep{olabi2022renewable,hoegh2019human}. However, direct utilization of electrical energy from renewable sources, such as wind and solar power, poses problems due to their intermittent nature~\citep{kousksou2014energy}. Moreover, storing the excess renewable electricity during high production periods is non-trivial. While pumped hydroelectric energy storage has provided the primary means for load balancing in an electrical grid in many countries~\citep{ferreira2013characterisation}, a wider portfolio of renewable electric energy storage solutions needs to be considered~\citep{gur2018review}. Alternatives include storage in batteries or conversion to hydrogen (H$_2$) via electrolysis~\citep{zhang2016survey}, for instance. Of these two, the latter was deemed more cost-efficient over longer storage times in a recent screening financial analysis~\citep{penev2019energy}, rendering it a potential candidate as an energy storage medium. Hydrogen may be directly utilized as a fuel or be further processed to {\it{e.g.}} liquid hydrocarbon and alcohol fuels. These synthetic fuels can then be employed in transport applications which are challenging to directly electrify~\citep{dieterich2020power}.

In this context, methanol (CH$_3$OH) is a suitable candidate for serving either as a fuel or as a feedstock for deriving other fuels and products~\citep{dieterich2020power}. Methanol can be chemically synthesized by hydrogenation of \ce{CO2} and \ce{CO}. To date, methanol is produced primarily from synthesis gas (syngas), a \ce{CO}-\ce{H2}-\ce{CO2}-mixture, which is derived from fossil fuels, especially natural gas and coal~\citep{bozzano2016efficient}. Direct hydrogenation of \ce{CO2} to methanol is also possible and has been demonstrated in practice~\citep{zhong2020state}. This could be a viable option to achieve carbon neutrality in methanol production, since in the future, both H$_2$ and CO$_2$ could be supplied on a sustainable basis instead of fossil sources. Indeed, H$_2$ can be supplied by biomass gasification~\citep{chang2011biomass} or water electrolysis where the required electricity is generated by renewables~\citep{oliveira2021green}, the former being the more mature technology at present. On the other hand, CO$_2$ could be captured from (industrial) flue gas sites or directly from the atmosphere using various carbon capture technologies~\citep{dieterich2020power,zhang2018effectiveness}.

Both syngas- and \ce{CO2}-based methanol synthesis is an exothermic process which requires a catalyst to drive the process, while thermodynamically the process prefers low temperatures and kinetically it favours higher temperatures~\citep{dieterich2020power}. Currently, the syngas processes by Johnson Matthey and Lurgi, utilizing fixed-bed reactor setups, are prevalent in commercial settings~\citep{bozzano2016efficient,mbatha2021power} and leverage economies of scale to reduce costs. The catalyst is typically copper-zinc(oxide) based~\citep{ali2015recent} and the reactors operate above 200$^\circ$C and at 50-100 bars~\citep{dieterich2020power}. The synthesis is highly equilibrium-limited, and reactant gases have to be recycled into the reactor over multiple passes~\citep{tonkovich2008methanol,bakhtiary2020modelling}. Additionally, efficient heat removal is a crucial aspect of reactor design~\citep{dieterich2020power}. To this end, other reactor designs, including microstructured reactors, could prove beneficial in portable and small-scale applications, such as \ce{CO2}-based synthesis, due to their excellent heat and mass transfer properties~\citep{izbassarov2022three,an2012computational}. Some of these, such as honeycomb monolith reactors with longitudinal microchannels, also exhibit low pressure drop even at high gas flow rates~\citep{hosseini2020technological}, which typically incurs large pressure losses in packed-bed reactors. This renders them potentially lucrative in such scenarios~\citep{arab2014methanol}. Previously, microstructured setups for methanol synthesis have been discussed and experimentally studied in~\citep{tonkovich2008methanol,hayer2011synthesis}.

In reactor studies involving methanol synthesis, 1D plug-flow reactor models have been successfully applied to predict the reactor performance~\citep{bozzano2016efficient} assuming perfect radial mixing and steady-state conditions. Such 1D models can be further divided into pseudo-homogeneous and heterogeneous models~\citep{bozzano2016efficient}, with typically similar predictive capabilities at nominal reactor operating conditions~\citep{manenti2011considerations}. 2D extensions of these models have also been proposed: in~\citep{arab2014methanol}, the authors compared the mass and heat transfer performance of packed-bed and monolithic reactors operating at identical conditions at industrial level, while in~\citep{petera2013new}, a two-dimensional model was utilized to describe the temporal evolution of temperature hotspots due to local decrease of catalyst diameter and bed porosity. \cite{eksiri2020two} studied an axial-radial flow pattern in a novel plate reactor design. In~\citep{solsvik2013multicomponent}, the authors applied a 2D model to analyze several closures for the mass diffusion fluxes at the pellet and reactor levels. In the study, the Wilke model provided similar predictive accuracy to the more rigorous Stefan-Boltzmann description of the intra-pellet mass diffusion.  

At the expense of increased computational effort, methanol synthesis can also be modeled by the computational fluid dynamics (CFD) approach, where the full set of Navier-Stokes equations are augmented with the necessary chemical kinetic models. The solid catalyst is then included typically via porous media models (PMM) and the particle-resolved computational fluid dynamics (PRCFD)~\citep{jurtz2019advances,dixon2020computational,micale2022computational}. The former treats the solid catalyst phase as a continuum and the pressure loss due to the catalyst is accounted for by source terms in the momentum equation. The latter is more realistic as it incorporates the catalyst explicitly and a fluid-solid interface is therefore present~\citep{micale2022computational}. In contrast to PMM modeling, this allows for local flow phenomena around the catalyst surface, such as secondary flows, vortex formation and flow separation to be captured~\citep{dixon2020computational}, which may be desired if their contribution to heat and mass transfer is critical~\citep{jurtz2019advances}. Consequently, this can result in superior modeling accuracy in the presence of more complex reactor geometries and turbulence, for instance. However, such studies are scarce at present with respect to methanol synthesis. Mirvakili {\it{et al.}} studied the effects of flow mal-distribution in a Lurgi reactor~\citep{mirvakili2018cfd} and Redondo {\it{et al.}} investigated various tubular reactor designs~\citep{redondo2019intensified}. Karthik {\it{et al.}} studied the impact of catalyst shape on the reaction characteristics~\citep{karthik2020computational} while Izbassarov {\it{et al.}} studied the performance of a novel modular millireactor design and the impact of the geometric configuration of the catalyst on methanol yield in a fixed-bed reactor~\citep{izbassarov2021numerical,izbassarov2022three}. Additionally, \cite{kyrimis2021understanding} studied the role of \ce{CO} hydrogenation in a fixed-bed reactor.

While improved catalyst and reactor designs are highlighted for enabling a further penetration of synthetic methanol into the global energy market~\citep{dieterich2020power,rahimpour2008two}, little computational work has been conducted on synthetic methanol production in microchannel reactors, despite their appealing properties~\citep{sharifianjazi2023methane} specifically at high gas flow. Therefore, this study aims to address this research gap by presenting reactive square duct flow simulations of synthetic methanol production in low (laminar) and high (turbulent) gas flow. Two laminar and two turbulent flow cases are studied by direct numerical simulation (DNS). Such a computationally intensive modeling choice constrains the study to a single duct, representative of {\it{e.g.}} a single channel in a monolith. However, DNS avoids usage of additional turbulence models. The paper is organized as follows: first, the theoretical and numerical framework along with the relevant modeling assumptions is presented. Second, the setup, mesh and boundary conditions are detailed and third, the predicted reaction performance and turbulence-chemistry interactions are presented and analyzed. Finally, the conclusions and implications of the study as well as prospects for future work are discussed.  

\section{Theory and methods}

\subsection{The Navier-Stokes equations and the discretization framework}

Here, the fluid flow is considered to be modestly compressible at the low Mach number limit. The flow dynamics is governed by the Navier-Stokes equations:
\begin{align}
    & \frac{\partial \rho}{\partial t} + \nabla \cdot \left( \rho\mathbf{u} \right) = 0 \label{eq:NS1}, \\
    & \frac{\partial \rho\mathbf{u}}{\partial t} + \nabla \cdot \left(\rho\mathbf{u} \mathbf{u} \right) =
    -\nabla p + \nabla \cdot \left[\mu \left(\nabla\mathbf{u} + \nabla\mathbf{u}^T \right) - \frac{2}{3}\mu \nabla \cdot \mathbf{u}\mathbf{I}\right] \label{eq:NS2} \\
    & \frac{\partial \rho Y_k}{\partial t} + \nabla \cdot \left(\rho \mathbf{u} Y_k \right) = \nabla \cdot \left(\rho D \nabla Y_k \right) + \dot{\omega}_k  \tag{$k=1,\ldots,n_s$} \label{eq:NSSpecies} \\
    & \frac{\partial \rho h_t}{\partial t} + \nabla \cdot \left(\rho \mathbf{u} h_t \right) = \frac{\partial p}{\partial t} + \nabla \cdot \left(\frac{\lambda}{c_p} \nabla h_s \right) + \dot{\omega}_h \label{eq:NS3} .
\end{align}
The most important assumptions employed in this context are as follows
\begin{itemize}
    \item The ideal gas law is employed for density: $\rho = p \cdot \mathrm{MW}_t / \left(RT\right)$.
    \item The unity Lewis number approximation is used, {\it{i.e.}} $D = \lambda / \left( \rho c_p \right)$.
    \item The total enthalpy $h_t$ is defined as $h_t = h_s + (1/2)|\mathbf{u}|^2$.
\end{itemize}
The remaining symbols with their respective definitions are detailed in Table~\ref{tab:tab1}.
\begin{table}[t!]
    \centering
    \begin{tabular}{l|c|r} 
         Symbol & Property & Unit (SI)\\ \hline
         $\alpha_{jk}$ & stoichiometric coefficient of specie $k$ & -- \\
         $A_i$ & geometric area of catalytic wall at face $i$ & m$^2$\\
         $c_p$ & heat capacity at constant pressure & J kg$^{-1}$ K$^{-1}$ \\
         $D$ & mass diffusivity & m$^2$ s$^{-1}$ \\ 
         $h_{f,k}^0$ & enthalpy of formation for specie $k$ & J/kg \\
         $h_t$ & total enthalpy & J/kg \\ 
         $h_s$ & sensible enthalpy & J/kg \\
         $I$ & identity tensor & --\\
         $\lambda$ & thermal conductivity & W K$^{-1}$ m$^{-1}$ \\ 
         MW & molecular weight & g/mol \\
         MW$_k$ & molecular weight of specie $k$ & g/mol \\
         MW$_t$ & total molecular weight of a gas mixture & g/mol \\
         $\mu$ & (dynamic) viscosity & Pa $\cdot$ s \\
         $n_s$ & number of species & -- \\
         $n_r$ & number of reactions & --\\
         $n_f$ & number of faces & --\\
         $\dot{\omega}_k$ & chemical reaction rate for specie $k$ & kg m$^{-3}$ s$^{-1}$ \\
         $\dot{\omega}_h$ & heat release rate & kg m$^{-1}$ s$^{-3}$ \\
         p & pressure & Pa \\
         $r_j$ & reaction rate of reaction $j$ & kmol kg$^{-1}_\mathrm{cat}$ s$^{-1}$ \\
         $\rho$ & density & kg m$^{-3}$ \\
         $\rho_c$ & (surface) density of catalyst & kg$_\mathrm{cat}$ m$^{-2}$ \\
         $t$ & time & s\\
         $T$ & temperature & K \\
         $\mathbf{u}$ & velocity & m s$^{-1}$ \\
         $Y_k$ & mass fraction of specie $k$ & -- \\
    \end{tabular}
    \caption{The description of symbols as they appear in the study.}
    \label{tab:tab1}
\end{table}

In the equations above, the chemical reaction rate of specie $k$ ($\dot{\omega}_k$) and heat release rate ($\dot{\omega}_h$) need to be further defined. In the context of methanol synthesis, the following relation has been introduced and successfully utilized in the previous work by Izbassarov {\it{et al.}}~\citep{izbassarov2021numerical,izbassarov2022three}
\begin{equation}
    \dot{\omega}_k = \rho_{c} \mathrm{MW}_k \sum_{i=1}^{n_f} \sum_{j=1}^{n_r} \frac{A_i \alpha_{jk} r_{ij}}{V_i} \label{eq:NSSpecieSource} .
\end{equation}
We refer the reader to the work cited above for further details. The heat release is then related to this reaction rate of specie $k$
\begin{equation}
    \dot{\omega}_h = \sum_{i=1}^{n_s} \Delta h_{f,k}^0 \dot{\omega}_k .
\end{equation}

The system of equations introduced above is numerically discretized using the finite volume method (FVM) code OpenFOAM (Foundation version 7), an open-source CFD software suite~\citep{weller1998tensorial}. The PIMPLE algorithm for pressure is used utilizing 3 outer corrector and 4 inner corrector steps. Furthermore, a first-order Euler implicit scheme is used for temporal discretization while the central differencing is applied for spatial discretization of diffusion terms. For the convective terms, the second-order accurate Gamma scheme~\citep{jasak1999high} with a blending factor of 0.05 is employed (see~\ref{appx:appx1}). 

Similar to the previous studies by Izbassarov {\it{et al.}}~\citep{izbassarov2021numerical,izbassarov2022three}, supporting simulations in Aspen Plus with the (0D) Gibbs reactor model neglecting reaction kinetics are performed. The equilibrium methanol yield derived from the Aspen Plus result provides the upper limit benchmark value for the local methanol yield at the catalytic walls in the full 3D simulations.

\subsection{The chemical kinetic model} \label{sec:chemistrymodel}
Here, we utilize the reaction mechanism and chemistry solver as previously used and validated by Izbassarov~{\it{et al.}}~\citep{izbassarov2021numerical,izbassarov2022three}. Next, the main features of the model are briefly summarized. The kinetic model by Graaf {\it{et al.}}~\citep{graaf1988kinetics}, updated with the equilibrium constants by Lim~\citep{lim2009modeling} and experimental data provided by An {\it{et al.}}~\citep{xin2009methanol}, is utilized here as presented in~\citep{kiss2016novel}. The model was designed for Cu/Zn/Al/Zr catalyst and considers the three primary reactions
\begin{flalign}
    && (\mathrm{R}_1) && \ce{CO + 2H2 <=> CH3OH} && \ce{\Delta H_{298K} = -90.2 kJ/mol} \nonumber \\
    && (\mathrm{R}_2) && \ce{CO2 + H2 <=> CO + H2O} && \ce{\Delta H_{298K} = +41.3 kJ/mol} \nonumber \\
    && (\mathrm{R}_3) && \ce{CO2 + 3H2 <=> CH3OH + H2O} && \ce{\Delta H_{298K} = -48.8 kJ/mol} \nonumber ,
\end{flalign}
where reactions 1 and 3 result in methanol forming via hydrogenation of \ce{CO} and \ce{CO2}, respectively, while reaction 2 is the reverse water-gas shift reaction (RWGSR) mediating the concentrations of CO$_2$ and CO in the presence of water. The model is reduced based on rate-determining steps from the full set of intermediate reactions considering Langmuir-Hinshelwood-Hougen-Watson (LHHW) kinetics involving 2 reactive sites at the catalyst. Despite this reduction and subsequently simpler kinetic description, the model predictions compare favorably with experimental data~\citep{xin2009methanol} and is therefore utilized in the present study.

Following the presentation of Kiss {\it{et al.}}~\citep{kiss2016novel}, the chemical reaction rate $r_j$ reads
\begin{align}
    & r = \frac{A B}{C} , \\
    & A = a T^n \exp\left( \frac{-E_a}{R T} \right), \\
    & B = B_1 \left( \prod f_i^{b_i} \right) - B_2 \left( \prod f_j^{b_j} \right), \\
    & C = \left( C_i \left[ \prod f_j^{c_j} \right]\right)^m, \\
\end{align}
where the terms $A$, $B$ and $C$ refer to the kinetic term, driving force and adsorption term, respectively.
Finally, the respective reaction rates for reactions 1, 2 and 3 read
\begin{align}
    & r_{1} = k_{1} K_{CO} \left(f_{CO} f_{H_2}^{3/2} - f_{CH_3OH}/[f_{H_2}^{1/2} K_{eq_{1}}] \right) \chi^{-1}, \\
    & r_{2} = k_{2} K_{CO_2} \left(f_{CO_2} f_{H_2} - f_{H_2O}f_{CO} /[K_{eq_{2}}] \right) \chi^{-1}, \\
    & r_{3} = k_{3} K_{CO_2} \left(f_{CO_2} f_{H_2}^{3/2} - f_{CH_3OH} f_{H_2O}/[f_{H_2}^{3/2} K_{eq_{3}}] \right) \chi^{-1}, \\
    & \chi = \left( 1 + K_{CO} f_{CO} + K_{CO_2}f_{CO_2} \right) \left[f_H^{1/2} + \left(K_{H_2O}/K_{H_2}^{1/2} \right) f_{H_2O}\right] ,
\end{align}
where the fugacity $f$ is approximated using the ideal gas law and corresponds to the partial pressure of the gas. The remaining coefficients are described in Table~\ref{tab:tab2}. The rates $r_{1}$, $r_{2}$ and $r_{3}$ introduced above can now be incorporated into Eq.~\eqref{eq:NSSpecieSource} and close the system of equations with appropriate boundary conditions (BC). Here, the chemistry is explicitly solved without operator splitting.
\begin{table}[t!]
    \centering
    \begin{tabular}{l|c|c|r} 
         Symbol & Expression & Unit (SI)\\ \hline
         $k_{1}$ & 4.0638 $\times$ $10^{-6}$ $\exp\left( -11695/ R T\right)$ & kmol kg$_{\mathrm{c}}^{-1}$ s$^{-1}$ Pa$^{-1}$ \\
         $k_{2}$ & 9.0421 $\times$ $10^{8}$ $\exp\left( -112860/ R T\right)$ & kmol kg$_{\mathrm{c}}^{-1}$ s$^{-1}$ Pa$^{-1/2}$ \\
         $k_{3}$ & 1.5188 $\times$ $10^{-33}$ $\exp\left( -266010/ R T\right)$ & kmol kg$_{\mathrm{c}}^{-1}$ s$^{-1}$ Pa$^{-1}$ \\
         $K_{CO}$ & 8.3965 $\times$ $10^{-11}$ $\exp\left( 118270/ R T\right)$ & Pa$^{-1}$ \\
         $K_{CO_2}$ & 1.7214  $\times$ $10^{-10}$ $\exp\left( 81287/ R T\right)$ & Pa$^{-1}$ \\
         $K_{CO} / K_{eq_{1}}$ & 3.5408 $\times$ $10^{12}$ $\exp\left( 19832/ R T\right)$ & Pa \\
         $K_{CO_2} / K_{eq_{2}}$ & 6.1221 $\times$ $10^{-13}$ $\exp\left( 125226/ R T\right)$ & Pa$^{-1}$ \\
         $K_{CO_2} / K_{eq_{3}}$ & 2.5813 $\times$ $10^{10}$ $\exp\left( 26788/ R T\right)$ & Pa \\
         $K_{H_2O} / K_{H_2}^{1/2}$ & 4.3676 $\times$ $10^{-12}$ $\exp\left( 115080/ R T\right)$ & Pa$^{-1/2}$ \\
    \end{tabular}
    \caption{The coefficients for reactions 1, 2 and 3.}
    \label{tab:tab2}
\end{table}

The important modeling assumptions regarding the chemical reactions are as follows
\begin{itemize}
    \item Only surface reactions are considered. The active surface area of the catalyst is its geometric area ($A_i$ in Eq.~\eqref{eq:NSSpecieSource}).
    \item Diffusion of reactants within the catalyst is neglected.
    \item Catalyst deactivation is negligible and not modeled.
    \item Side-reactions are omitted (such as conversion to dimethyl ether).
\end{itemize}
Similar assumptions are common in the literature and are consistent with the assumptions in recent CFD studies of synthetic methanol production involving microreactor designs~\citep{izbassarov2021numerical,izbassarov2022three}. 

\subsection{Numerical setup} \label{sec:numsetup}
\begin{figure}[ht!]
\centering
 \includegraphics[width=0.98\textwidth]{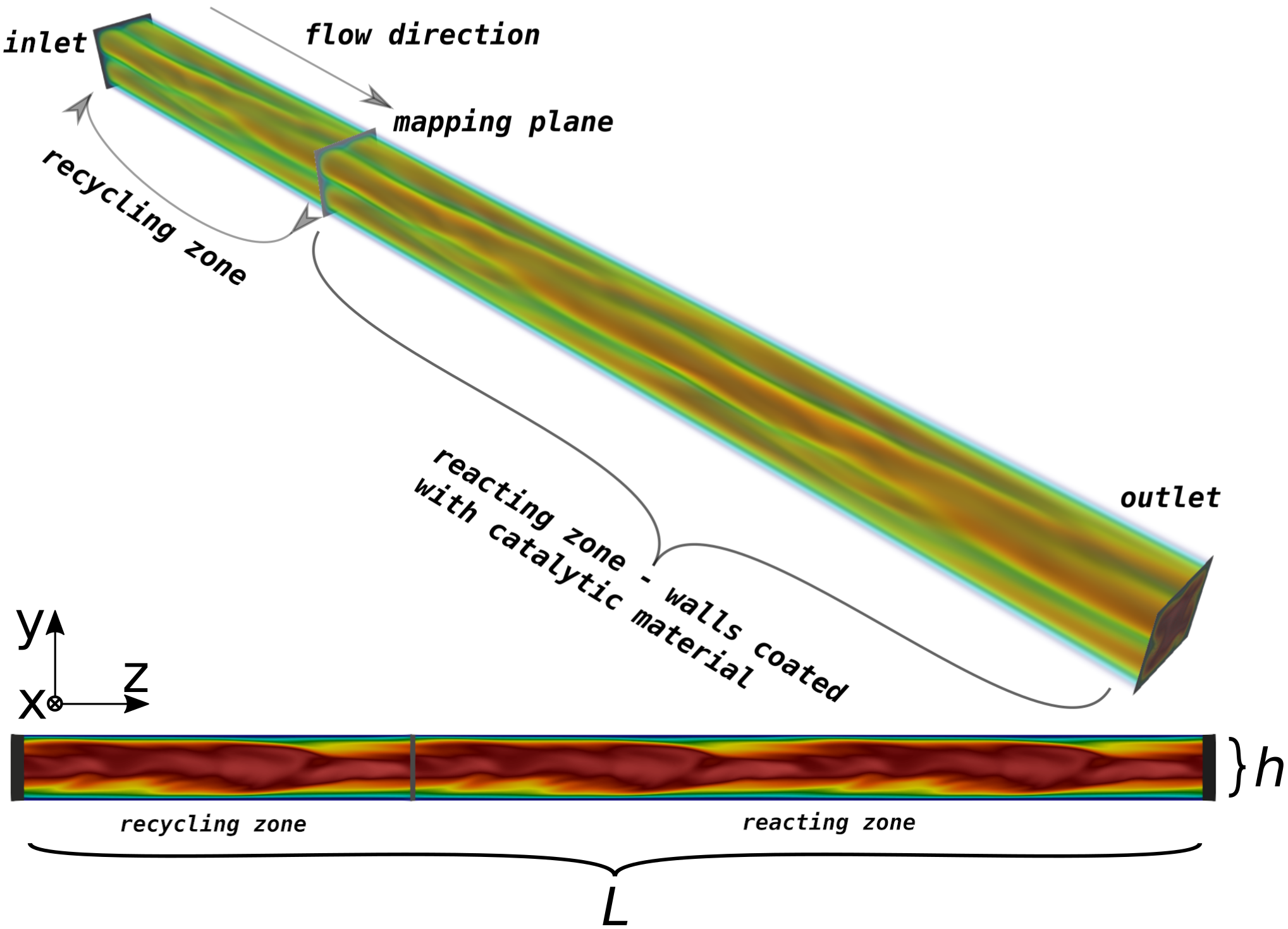}
 \caption{An isometric view of the complete reactive duct (top) and a sideview of the duct center (bottom). In both figures, the magnitude of the instantaneous velocity in a turbulent flow ($Re_{\mathrm{b}} = 1100$) is depicted with green (red) indicating regions of low (high) velocity. To generate a turbulent flow with the correct statistics at the inlet, a strong recycling approach is used: the reactive region of the duct is separated from the recycling zone, where the turbulent flow is generated and recirculated by applying periodic boundary conditions on the inlet and the mapping plane.}
 \label{fig:fig1}
\end{figure}
In the present study, we investigate laminar and turbulent flow at a fixed bulk Reynolds number in a square duct with catalytic walls, as illustrated in Fig.~\ref{fig:fig1}. Notably, the width and height ($h$) of the square duct are set to 2 mm, comparable to the single-channel cross-sectional dimensions in typical monolithic reactors~\citep{arab2014methanol}. The flow proceeds from the inlet towards the outlet, and the duct consists of two regions, the recycling zone (no wall reactions) and the reacting zone (wall reactions), the former being necessary to ensure fully developed flow as the reactants enter the reacting zone. In the reacting zone, methanol synthesis occurs at the walls, which are coated with the catalyst material and are referred to as the catalytic walls hereafter. The strong recycling scheme (see~\citep{wu2017inflow}) is employed to develop the flow within the recycling zone, {\it{i.e.}} the velocity field from the downstream boundary (mapping plane in Fig.~\ref{fig:fig1}) of the recycling zone is recycled back to the inlet. Subsequently, the streamwise velocity component is scaled by a constant to ensure a volumetric mass flux consistent with the set bulk Reynolds number. The latter is defined here as $Re_\mathrm{b} = \rho U_\mathrm{ave} h / \left( 2 \mu \right)$, where $\rho$ is the gas density, $U_\mathrm{ave}$ is the surface-averaged streamwise velocity at the inlet, $h/2$ is the duct half-width and $\mu$ is the dynamic viscosity. Each simulation case is run in two steps: the first step involves a non-reactive flow to fully develop the flow to correct statistics, and subsequently, a second (reactive flow) run is performed with methanol synthesis enabled at the catalytic walls. To attain a fully developed flow, a plug-flow profile is first imposed in the recycling region, and a non-reactive flow simulation is conducted for 10 passes with the recycling enabled (laminar cases). For the turbulent cases, random noise is also added at the onset of the simulation to trigger turbulence in the flow. In these turbulent cases ($Re_\mathrm{b}$=1100 and 2200), additional runs are also performed without adding the noise to yield a non-turbulent, laminar flow as references for assessing the influence of turbulent fluctuations on the reaction characteristics. Such a situation could occur e.g. in a long duct with smooth walls so that the transition to turbulence is delayed. Notably, the strong recycling scheme is considered superior to most present alternatives ({\it{e.g.}} synthetic methods)~\citep{wu2017inflow} and a similar approach has been adopted in several studies successfully~\citep{lucci2013three,arani2017three,arani2018direct,arani2019direct}.

After this initialization, the full reactive flow can be simulated with the appropriate BCs summarized in Table~\ref{tab:tab3_1}. Notably, the setup is isothermal to exclude the effects of heat transfer on the reaction performance and possible flow laminarization in the turbulent cases. The isothermal configuration has also been consistently found to possess superior conversion performance to adiabatic reactors in similar studies~\citep{izbassarov2021numerical,izbassarov2022three}. The temperature window for optimal methanol synthesis typically ranges from 500 to 525 K~\citep{kiss2016novel,izbassarov2021numerical,izbassarov2022three}, and thus, a fixed temperature value $T=503.15$ K is set at the inlet and each duct wall here. The pressure inside the duct is maintained at $p=50$ bar, which is a common operating pressure in commercial reactors. Further, the reactants introduced at the inlet are \ce{H2} and \ce{CO2} with a molar ratio of R = [\ce{H2}]/[\ce{CO2}] = 3 (stoichiometric ratio). A high catalyst mass $m_c=24.5$ g is also deployed to produce low values of gas hourly space velocity (GHSV) even at high gas flow. This choice is assumed to promote transport-limited chemistry, where the potential impact of turbulence on methanol yield is elicited.
\begin{table}[t!]
    \centering
    \begin{tabular}{l|l|l|l|l} 
         Quantity & $p$ & $\mathbf{u}$ & $T$ & $m_c$ \\ \hline
         inlet & zero-gradient & mapped & 503 K & -- \\
         outlet & $5.0 \cdot 10^6$ Pa & zero-gradient & zero-gradient & -- \\
         walls (recyc. zone) & zero-gradient & no-slip & 503 K & -- \\
         walls (react. zone) & zero-gradient & no-slip & 503 K & 24.5 g\\
    \end{tabular}
    \caption{A summary of the boundary conditions in the present reactive duct flow simulations. The catalyst mass is denoted by $m_c$.}
    \label{tab:tab3_1}
\end{table}

The present solver and mesh have been validated for the highest Reynolds number ($Re_\mathrm{b}=2200$) turbulent duct flow DNS configuration of Zhang{~\it{et al.}}~\citep{zhang2015direct} presented in~\ref{appx:appx1}. For the reactive flow, identical mesh resolution is utilized and a reacting zone (recycling zone) length of 2.2 cm (1.2 cm) is considered. Further details regarding the mesh are presented in Table~\ref{tab:tab3}.
\begin{table}[t!]
    \centering
    \begin{tabular}{l|l|l} 
         Symbol & Property & Present simulations \\ \hline
         $N_x \times N_y \times N_z $ & Mesh dimensions & 128 $\times$ 128 $\times$ 792 \\
         $L$ & Length [recycl. zone, entire duct] (m) & [0.014 0.036] \\
         $h$ & Duct width/height (m) & 0.002 \\
         min $\Delta y^+$ & min. scaled wall unit & 0.48 \\
         $\gamma$ & Exp. ratio of adjacent wall-normal cells & 1.04 \\
         max $\Delta x^+ / \Delta y^+$ & Max. cell aspect ratio & 13.36 \\ 
    \end{tabular}
    \caption{A overview of the mesh properties in the present reactive duct flow simulations.}
    \label{tab:tab3}
\end{table}

\section{Results}
In the following, the main results derived from the DNS of the reactive flow involving synthetic methanol production in a square duct are presented and analyzed. The results are presented for four distinct (bulk) Reynolds numbers: 100, 500, 1100 and 2200. $Re_\mathrm{b}=100$ and $Re_\mathrm{b}=500$ yield fully laminar flow while $Re_\mathrm{b}=1100$ and $Re_\mathrm{b}=2200$ generate turbulent flow. Notably, while $Re_\mathrm{b} = 1100$ exceeds the minimum Reynolds number required for self-sustained turbulence in a square duct~\citep{owolabi2016experiments}, it produces only marginally turbulent flow. Specifically, we address the $Re_\mathrm{b}$-dependency of the methanol formation process at the catalytic walls focusing on the predicted conversion efficiency and measures of methanol yield. 

\subsection{General overview}
\begin{figure}[ht!]
\centering
 \includegraphics[width=0.98\textwidth]{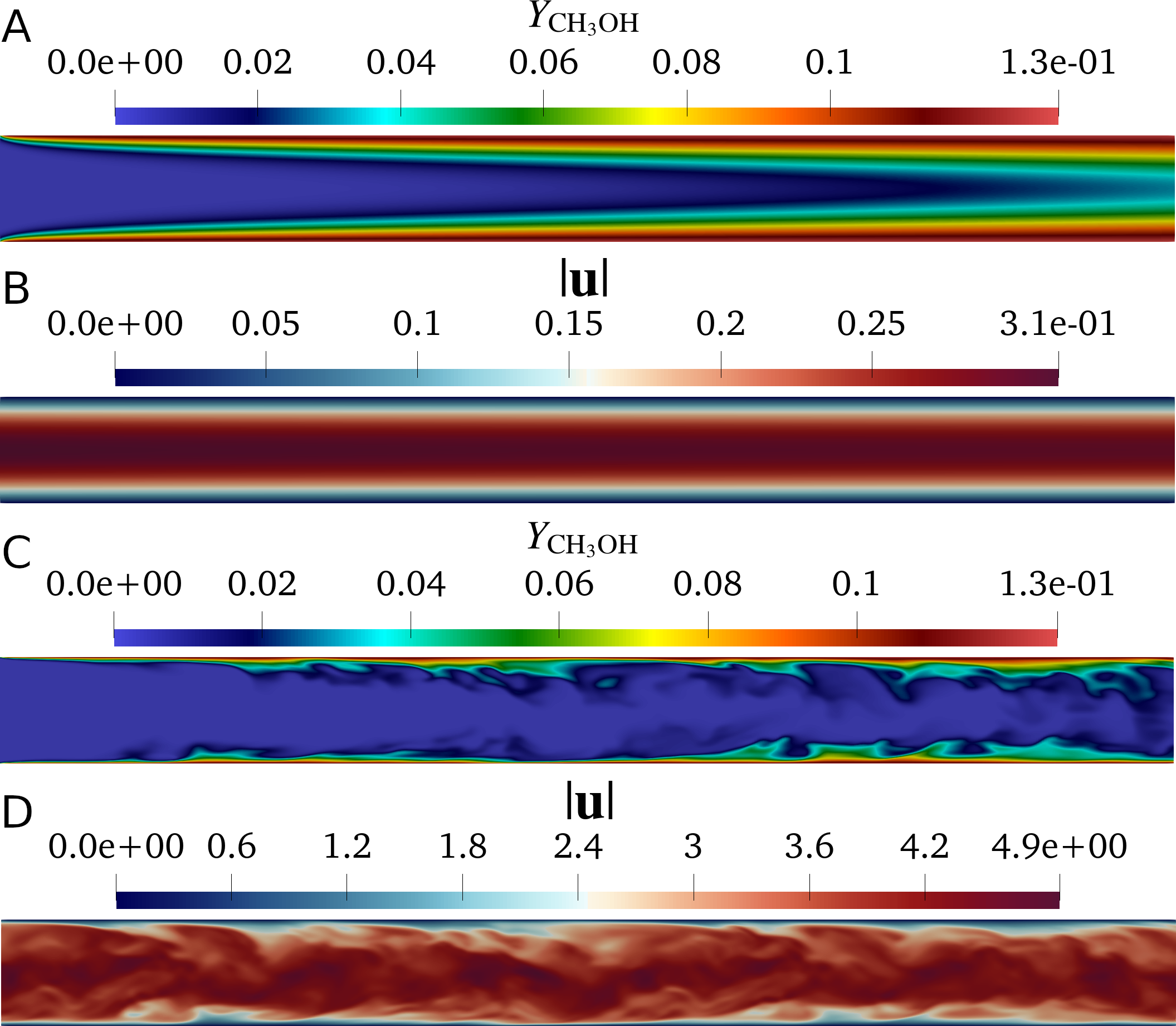}
 \caption{A, B: Midcuts of the reacting zone for the instantaneous methanol mass fraction and velocity magnitude at $Re_\mathrm{b}=100$, respectively. C, D: Midcuts of the reacting zone for the instantaneous methanol mass fraction and velocity magnitude at $Re_\mathrm{b}=2200$, respectively.}
 \label{fig:fig2}
\end{figure}
Fig.~\ref{fig:fig2} displays midcuts at the duct centerline of the (instantaneous) methanol mass fraction $Y_\mathrm{\ce{CH3OH}}$ as well as the velocity magnitude $|\mathbf{u}|$ for the laminar $Re_\mathrm{b}=100$ (Fig.~\ref{fig:fig2} A, B) and turbulent $Re_\mathrm{b}=2200$ case (Fig.~\ref{fig:fig2} C, D), respectively. Only the reacting zone is shown herein and the flow propagates from left to right. As the surface reactions only occur on the catalytic walls, \ce{CH3OH} is generated on the walls forming laminar (Fig.~\ref{fig:fig2} A) and turbulent (Fig.~\ref{fig:fig2} C) methanol mass transfer boundary layers which increase in thickness downstream as expected. In contrast, the velocity field in both cases is fully developed in the reacting zone due to the described flow recycling procedure. For the turbulent case, the isothermal setup ensures that flow does not laminarize downstream. Such a laminarization could occur if the gas at the catalytic walls is heated, resulting in increased local gas viscosity. Indeed, the turbulent statistics remain unaltered as a maximum attenuation of 2-3 \% is observed in the Reynold stress components along the length of the duct, the minor effect being caused by the increased density of the reaction products. 

\begin{figure}[ht!]
\centering
 \includegraphics[width=0.98\textwidth]{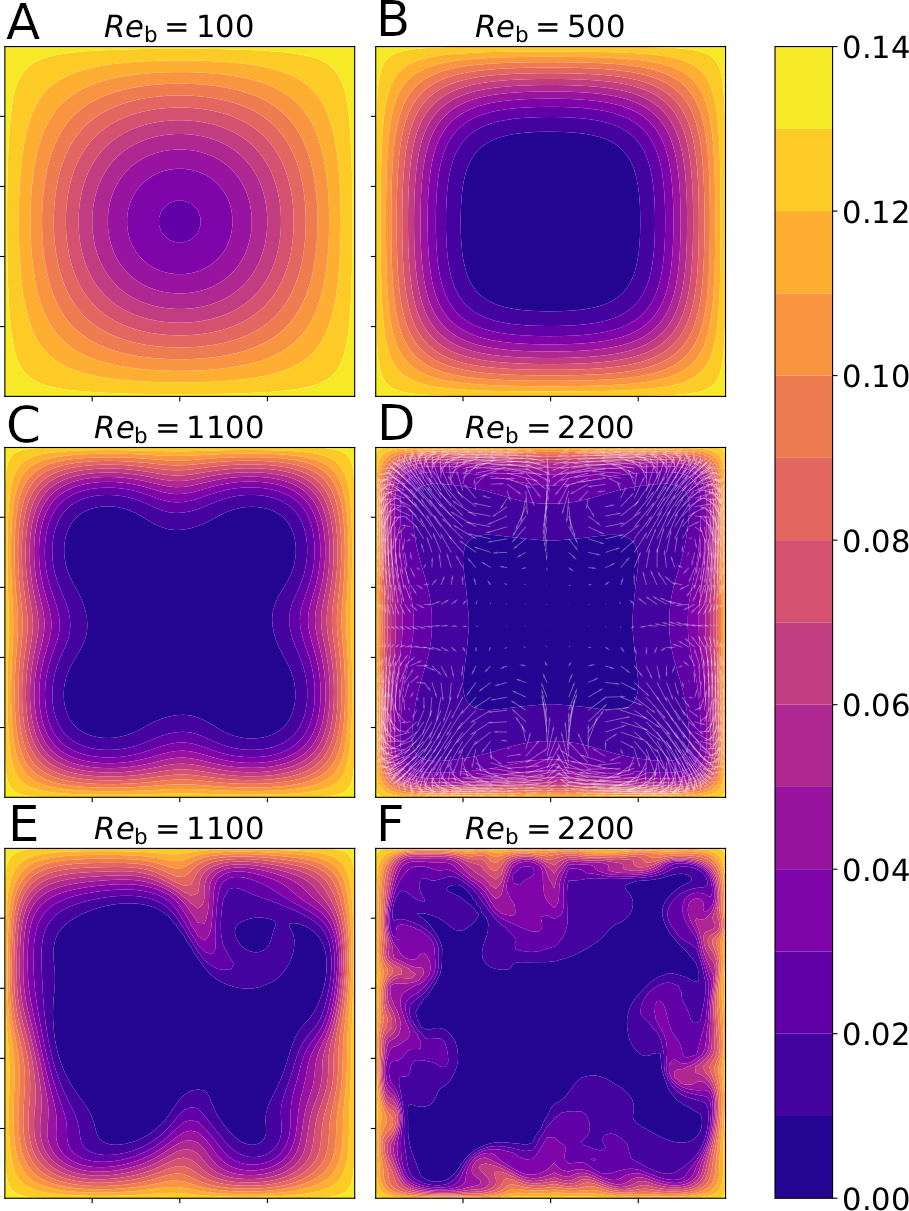}
 \caption{A-D: the temporally averaged methanol mass fraction at the outlet for the 4 simulation cases $Re_\mathrm{b}=100$, $Re_\mathrm{b}=500$, $Re_\mathrm{b}=1100$ and $Re_\mathrm{b}=2200$. E-F: the instantaneous methanol mass fraction at the outlet for $Re_\mathrm{b}=1100$ and $Re_\mathrm{b}=2200$ is also presented.}
 \label{fig:fig3}
\end{figure}
Furthermore, the cross-sectional methanol mass fraction profiles at the duct outlet are also presented in Fig.~\ref{fig:fig3}. The first and second rows display the time-averaged profiles for the laminar and turbulent cases, while the third row illustrates the instantaneous profiles for the turbulent flow regime. The mean flow patterns in the turbulent cases ($Re_\mathrm{b}=1100$ and $Re_\mathrm{b}=2200$) are reminiscent of non-reactive duct flow fields reported in the literature~\citep{gavrilakis1992numerical} and exhibit a distinct compression of the boundary layer towards the corners of the cross section not observed in the laminar cases ($Re_\mathrm{b}=100$ and $Re_\mathrm{b}=500$). This is due to the secondary flow patterns visible in the velocity vector fields at $Re_\mathrm{b}=2200$ (second row, right). Counter-rotating vortices forming at the duct corners induce recirculation and drive methanol towards the center of the walls, where the boundary layer is uplifted, observed as a bulging of the layer at these centers. The instantaneous mass fraction profiles (Fig.~\ref{fig:fig3} third row) also reveal the more subtle turbulent structures which become more pronounced as the Reynolds number is increased. The methanol boundary layer is elongated towards the duct center in regions where the turbulent velocity fluctuations involve the momentary ejection of low momentum fluid from the walls towards the center. Contrarily, the layer is compressed in regions with high momentum fluid surging from the duct center towards the walls, also known as sweep structures. The regions have a particular significance on the local chemical reaction rates which will be addressed later.

\subsection{Methanol yield}
Next, we discuss the methanol yield in the setup by several metrics. The averaged mass fraction profiles, reaction rates, mass flow rate and transfer as well as conversion efficiency are evaluated from the present DNS data and discussed both in the laminar and turbulent cases.

\begin{figure}[ht!]
\centering
 \includegraphics[width=0.98\textwidth]{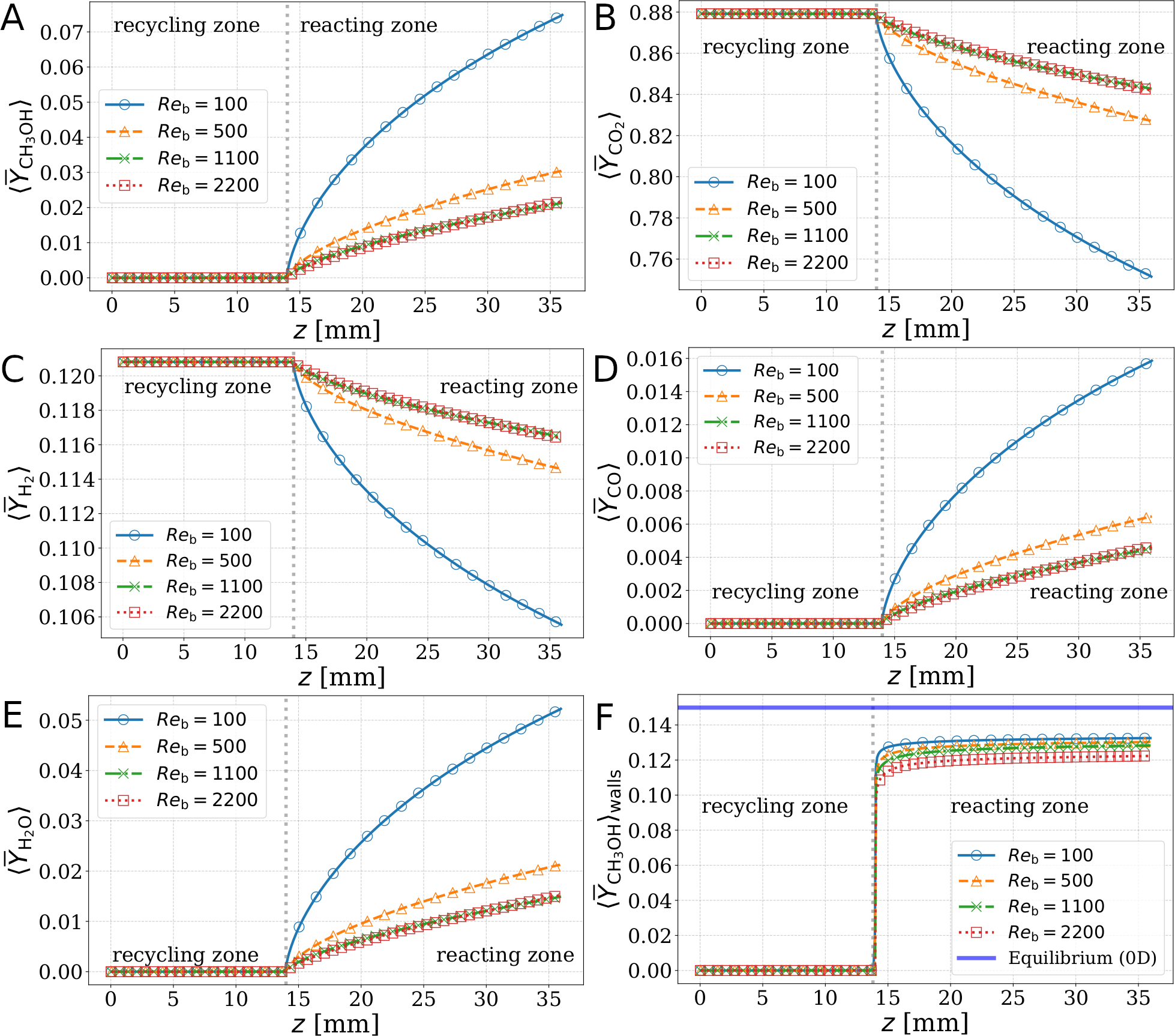}
 \caption{The time- and flux-averaged species mass fractions fields $\langle \overline{Y} \rangle$ presented over the streamwise coordinate $z$ for each simulation case. A-E: Profiles for \ce{CH3OH}, \ce{CO2}, \ce{H2}, \ce{CO} and \ce{H2O}, respectively. F: The time- and flux-average for \ce{CH3OH} mass fraction at the duct walls. The 0D equilibrium methanol yield attained in Aspen Plus with pressure and temperature conditions identical to the CFD cases studied here is also displayed.}
 \label{fig:fig4}
\end{figure}
Similar cross-sectional planes, as presented in Fig.~\ref{fig:fig3}, can also be defined across the length of the duct. At each plane, the time-averaged species mass fraction fields $\overline{Y} = \int Y dt / \int dt$ are surface-averaged with the local mass flux as the weighting factor: $\langle \overline{Y} \rangle = \int_A F \overline{Y} dA / \int F dA$, where $F$ is the mass flux. The resulting axial methanol profiles $\langle \overline{Y} \rangle$ are presented in Fig.~\ref{fig:fig4}. \ce{CH3OH}, \ce{CO} and \ce{H2O} (\ce{CO2} and \ce{H2}) mass fractions increase (reduce) monotonically upon entry from the recycling zone to the reacting zone ($z=14$ mm) as a function of the axial distance $z$ (Fig.~\ref{fig:fig4} A-E). Notably, the conversion rate to methanol is reduced as $Re_\mathrm{b}$ becomes higher, since an increased gas inflow implies lower residence time for the reactants at the catalytic walls. However, in the turbulent regime, $Re_\mathrm{b}=2200$ produces a higher conversion to products compared to $Re_\mathrm{b}=1100$, which is likely due to complementary effects of the fast chemistry and enhanced (turbulent) mixing, improving the transport of reactants to the catalytic walls. This argument is supported by Fig.~\ref{fig:fig4} F, where the flux-averaged methanol content at the catalytic walls is also shown. Here, the flux-averaging has been conducted exclusively on computational cells neighboring the catalytic walls, and therefore it embodies the average methanol wall content along the duct. The wall methanol mass fraction approaches a constant value very rapidly at each $Re_\mathrm{b}$ upon entry to the reacting zone due to the high catalyst loading. This indicates a fast, transport-limited chemistry in all the cases studied here. Fig.~\ref{fig:fig4} F also indicates the equilibrium-limited value predicted by the Aspen Plus 0D Gibbs reactor model, which disregards the reaction kinetics and is clearly an inaccurate description of the methanol synthesis in this setup.

\begin{figure}[ht!]
\centering
 \includegraphics[width=0.98\textwidth]{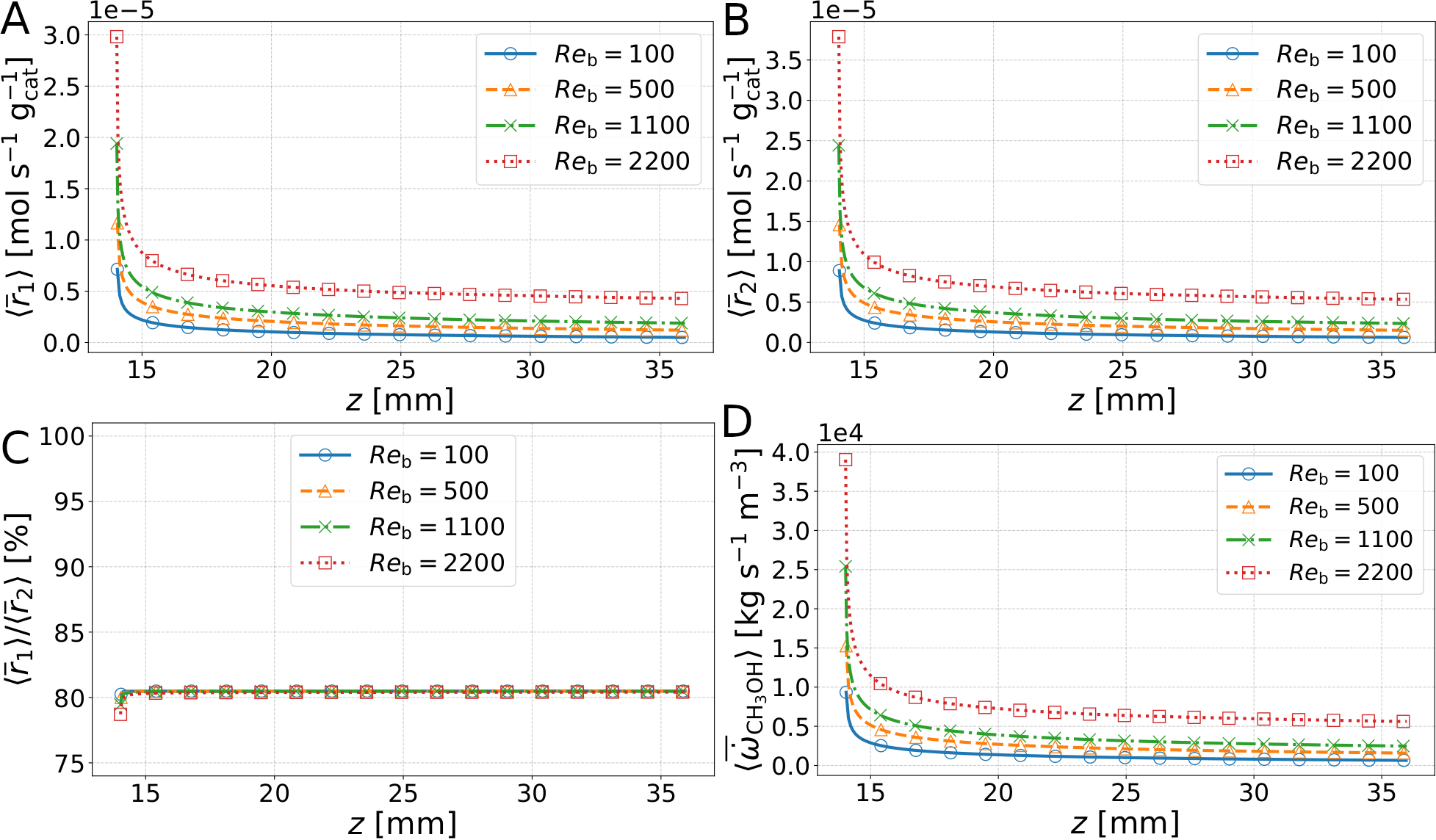}
 \caption{The time- and flux-averaged axial profiles of the reaction rates for A: reaction 1 and B: reaction 2. C: the relative fraction $\langle \overline{r}_1 \rangle$ / $\langle \overline{r}_2 \rangle$ of these two dominant reactions. D: the time- and flux-averaged net axial volumetric methanol formation rate $\langle \overline{\omega}_{\mathrm{CH}_3\mathrm{OH}} \rangle$ resulting from the individual reactions.}
 \label{fig:fig5}
\end{figure}

The time- and flux-averaged reaction rate terms for reaction 1 $\langle \overline{r}_\mathrm{1} \rangle$ and reaction 2 $\langle \overline{r}_\mathrm{2} \rangle$ are plotted in Fig.~\ref{fig:fig5} A and B, respectively.  The rates reach their maximum immediately upon entry to the reacting zone and then, as the catalytic surfaces become quickly saturated with methanol (Fig.~\ref{fig:fig4} F), decrease towards an asymptotic value, which is dictated by reactant transport limitations. Despite the lowered methanol mass fraction witnessed with increasing $Re_\mathrm{b}$ (Fig.~\ref{fig:fig4} A), the reaction rates increase in tandem with $Re_\mathrm{b}$, which then translates into a higher net volumetric methanol formation rate $\langle \overline{\dot{\omega}}_{\mathrm{CH}_3\mathrm{OH}} \rangle$ depicted in Fig.~\ref{fig:fig5} D. The impact of lower residence time discussed earlier is therefore compensated by the higher mass flux of reactants entering at higher Reynolds numbers. Importantly, the chemical equilibrium is not shifted from case to case since the temperature remains constant (isothermal setup) and the pressure drop across the duct is negligible, {\it{i.e.}} a near-constant pressure is maintained. Therefore, the relative magnitude $\langle \overline{r}_\mathrm{1} \rangle$/$\langle \overline{r}_\mathrm{2} \rangle$ of the two reactions remains constant as $Re_\mathrm{b}$ increases (Fig.~\ref{fig:fig5} C) and in particular, is unaffected by turbulence. 

As a remark, the kinetic model employed here predicts the conversion of \ce{CO2} to \ce{CO} (reaction 2 in Section~\ref{sec:chemistrymodel}) and a subsequent hydrogenation of \ce{CO} to methanol (reaction 1) as the primary route for methanol synthesis. To examine this with further simulations, the Aspen Plus 1D plug-flow reactor model as presented in~\citep{izbassarov2021numerical,izbassarov2022three} with matching reactor dimensions was deployed and the relevant range of operating pressure ($p=30-75$ bar), temperature ($T=180-300^\circ$ C) and reactant ratio $R=3-6$ was swept to study the rate of \ce{CO2} hydrogenation. Indeed, these simulations (not shown here) indicate that the corrected coefficients to the Graaf model based on the work of Lim{~\it{et al.}} and An{~\it{et al.}} as presented in~\citep{kiss2016novel} predict a negligible \ce{CO2} hydrogenation rate under these practical reactor operating conditions.  

\begin{figure}[ht!]
\centering
 \includegraphics[width=0.98\textwidth]{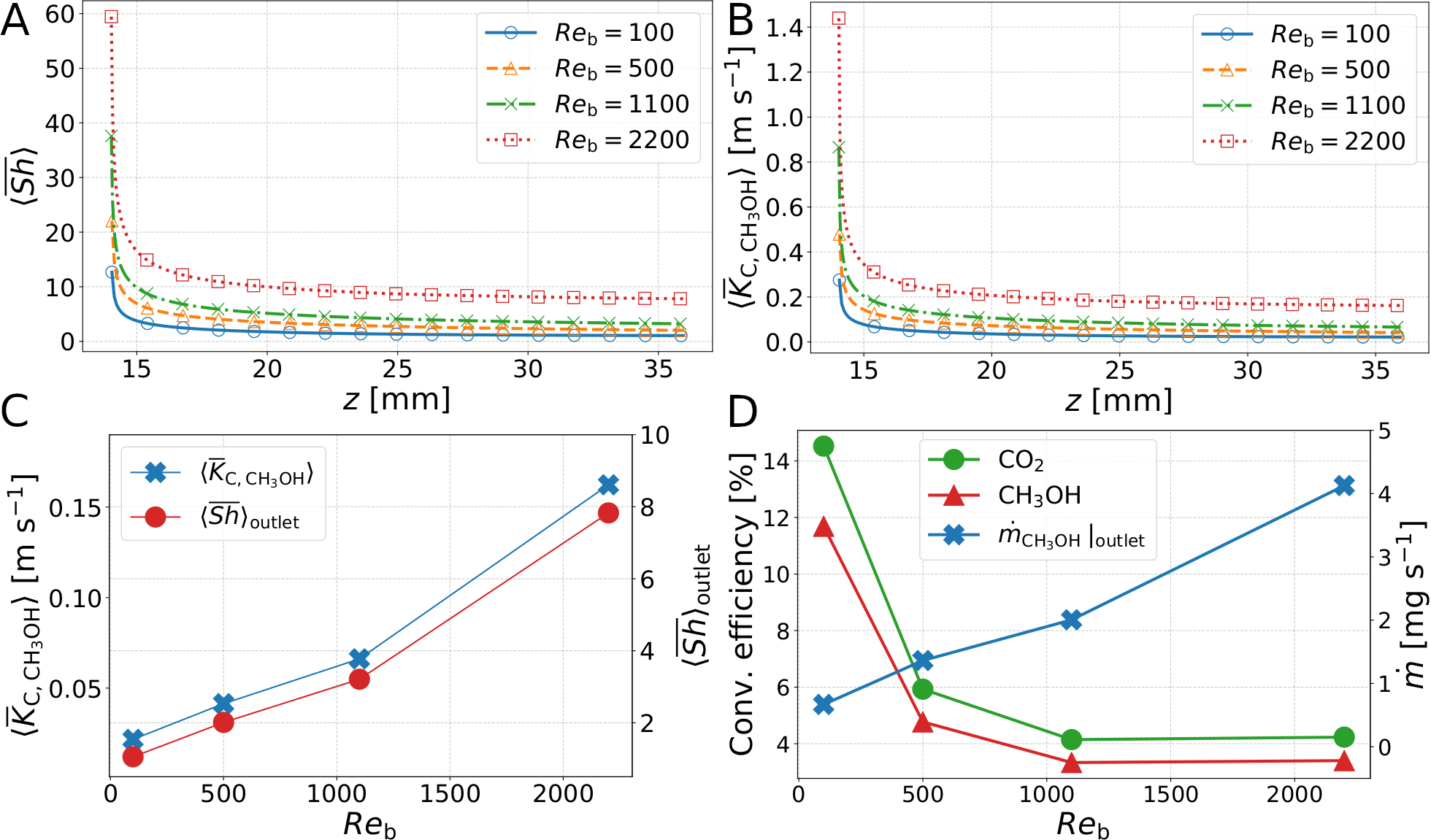}
 \caption{A: The axial variation of the averaged methanol Sherwood number $\langle \overline{Sh} \rangle$ at the catalytic walls. B: The axial evolution of the computed mass transfer coefficient $\langle \overline{K}_C \rangle$ of methanol evaluated at the catalytic walls. C: $\langle \overline{Sh} \rangle$ and $\langle \overline{K}_C \rangle$ values at the outlet. D: The conversion efficiency of \ce{CO2} and yield of \ce{CH3OH} as well as the methanol mass flow rate $\dot{m}$ at outlet for each $Re_\mathrm{b}$.}
 \label{fig:fig6}
\end{figure}
Fig.~\ref{fig:fig6} A illustrates the axial evolution of the averaged Sherwood number $\langle \overline{Sh}\rangle$ of methanol for each case, which is defined here as the non-dimensionalized methanol mass fraction gradient at the catalytic walls
\begin{equation}
    \langle \overline{Sh} \rangle = \frac{h}{2} \nabla \langle \overline{Y}_{\mathrm{CH}_3\mathrm{OH, walls}} \rangle / \left( \langle \overline{Y}_{\mathrm{CH}_3\mathrm{OH, walls}} \rangle - \langle \overline{Y}_{\mathrm{CH}_3\mathrm{OH,centerline}} \rangle \right), \nonumber
\end{equation}
where $h/2$ is the duct half-width and the denominator on the right-hand side describes the difference in the mean methanol mass fraction moving from the duct walls to the duct center. As depicted in Fig.~\ref{fig:fig2}, the methanol boundary layer is established at the onset of the reacting zone and gradually increases in thickness downstream, suggesting decreasing methanol mass fraction gradient. Thus, $\langle \overline{Sh} \rangle$ values peak and then tend towards the steady-state values, which are $\langle \overline{Sh} \rangle \approx 1.05$, 2.01, 3.21 and 7.83 for $Re_\mathrm{b}=100$, 500, 1100 and 2200, respectively and visualized in Fig.~\ref{fig:fig6} C. Similar axial plot for the mean methanol mass transfer coefficient $\langle \overline{K}_C \rangle$, defined as
\begin{equation}
    \langle \overline{K}_C \rangle = \frac{\langle \overline{Sh} \rangle D}{h/2} ,\nonumber
\end{equation}
is depicted in Fig.~\ref{fig:fig6} B. The functional form closely resembles that of $\langle \overline{Sh} \rangle$ since this is also the dominant term in the expression for $\langle \overline{K}_C \rangle$ and the outlet values for $\langle \overline{K}_C \rangle$ are visualized at each $Re_\mathrm{b}$ in Fig.~\ref{fig:fig6} C. The conversion efficiency of \ce{CO2}, the yield of \ce{CH3OH} and the mass flow rate of \ce{CH3OH} at the duct outlet as a function of $Re_\mathrm{b}$ are depicted in Fig.~\ref{fig:fig6} D. The conversion efficiency and yield are described~\citep{izbassarov2021numerical}
\begin{align}
\zeta_{\mathrm{CO}_2} &= \frac{X_{\mathrm{CO}_2\mathrm{,i}} - X_{\mathrm{CO}_2\mathrm{,o}}}{X_{\mathrm{CO}_2\mathrm{,i}}} = \frac{Y_{\mathrm{CO}_2\mathrm{,i}} - Y_{\mathrm{CO}_2\mathrm{,o}}}{Y_{\mathrm{CO}_2\mathrm{,i}}} , \nonumber \\
\zeta_{\mathrm{CH}_3\mathrm{OH}} &= \frac{X_{\mathrm{CH}_3\mathrm{OH}\mathrm{,o}}}{X_{\mathrm{CO}_2\mathrm{,i}}} = \frac{Y_{\mathrm{CH}_3\mathrm{OH}\mathrm{,o}} \mathrm{MW}_{\mathrm{CO}_2}}{Y_{\mathrm{CO}_2\mathrm{,i}} \mathrm{MW}_{\mathrm{CH}_3\mathrm{OH}}} \nonumber,
\end{align}
where the subscript $i$ ($o$) refers to mass fractions at the inlet (outlet). Notably, $\zeta_{\mathrm{CO}_2}$ is higher than $\zeta_{\mathrm{CH}_3\mathrm{OH}}$ for each Reynolds number since a degree of \ce{CO2} converted to \ce{CO} is not hydrogenated to \ce{CH3OH}. The maximum value for $\zeta_{\mathrm{CH}_3\mathrm{OH}}$ at $Re_\mathrm{b}=100$ is 11.7 \%, while $\zeta_{\mathrm{CO}_2} \approx$ 14.5 \%, which are relatively low in comparison to values reported in other microreactor studies~\citep{izbassarov2021numerical,izbassarov2022three}.
While the single-pass efficiency clearly deteriorates with increasing $Re_\mathrm{b}$, the mass flux of reactants also increases accordingly in the high Reynolds number cases as discussed earlier, resulting in higher volumetric methanol formation rates (Fig.~\ref{fig:fig5} D). Therefore, a higher net methanol mass flow rate $\dot{m}$ at the outlet is witnessed.

\subsection{Turbulence-chemistry interactions}
\begin{figure}[ht!]
\centering
 \includegraphics[width=0.49\textwidth]{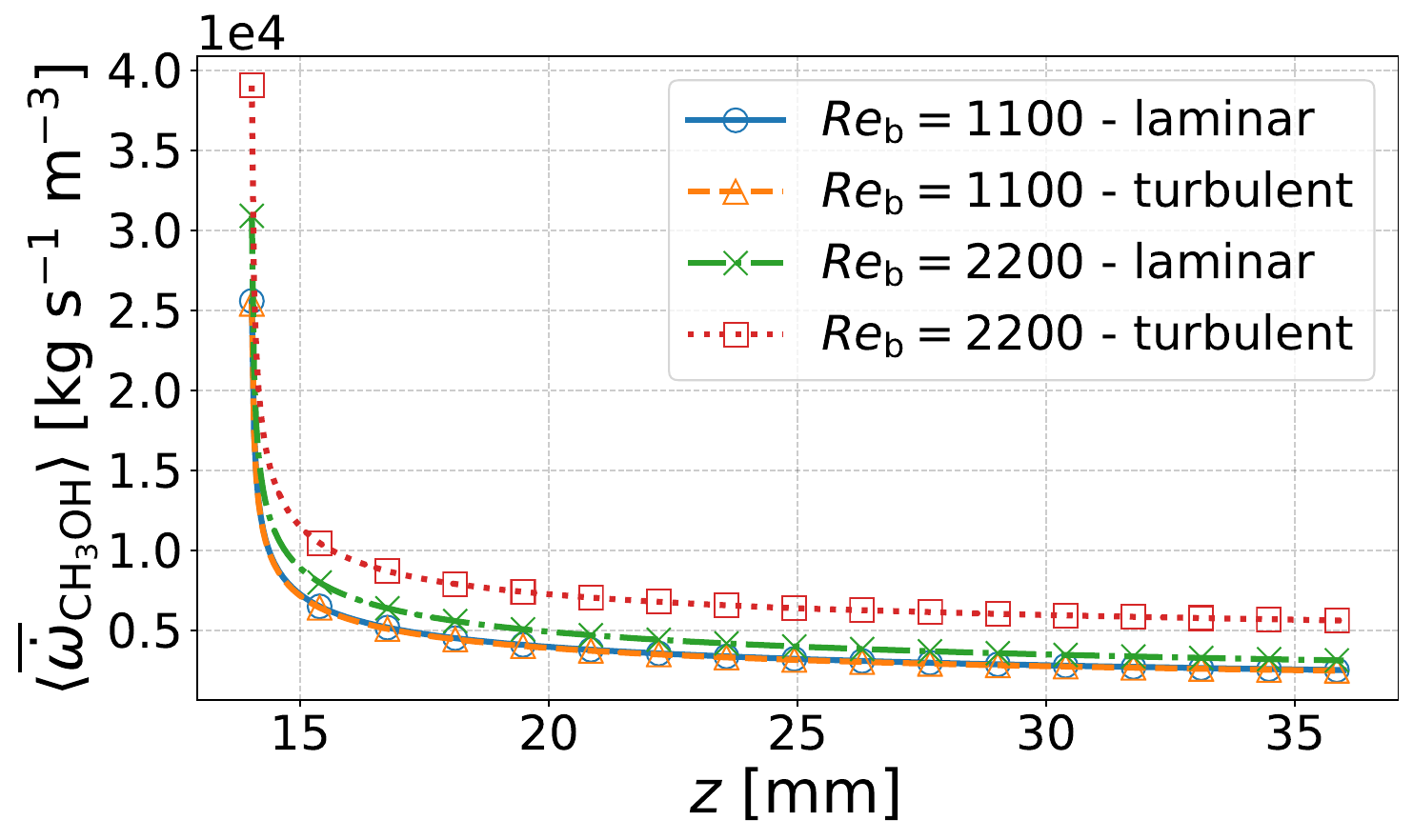}
 \caption{The axial volumetric methanol formation rate for the highest Reynolds number cases $Re_\mathrm{b}=1100$ and $Re_\mathrm{b}=2200$. To demonstrate the significance of turbulent transport on the formation rate, the values produced by non-turbulent laminar flow at corresponding $Re_\mathrm{b}$ are also plotted.}
 \label{fig:fig7}
\end{figure}
As described in Section~\ref{sec:numsetup}, the high Reynolds number cases ($Re_\mathrm{b}=1100$, $Re_\mathrm{b}=2200$) can be simulated in both laminar and turbulent modes. In the laminar modes, no turbulent structures develop in the flow as the initial random perturbations are disabled while initializing the flow. In the turbulent modes, these perturbations are included and physically correct turbulent flow is thus generated. Both modes at $Re_\mathrm{b}=1100$ and $Re_\mathrm{b}=2200$ are simulated to facilitate the analysis of the turbulent structures on methanol yield.
The differences in mean axial volumetric methanol formation rates $\langle \overline{\omega}_{\mathrm{CH}_3\mathrm{OH}} \rangle$ at the highest Reynolds numbers ($Re_\mathrm{b}=1100$, $Re_\mathrm{b}=2200$) under both turbulent and laminar flow conditions are compared in Fig.~\ref{fig:fig7}. The turbulent and laminar flow at $Re_\mathrm{b}=1100$ produce identical, overlapping methanol production profiles. This suggests that the turbulent fluctuations contribute little in terms of enhancing the yield at this marginally turbulent limit and will be thus excluded from the following discussion. However, for $Re_\mathrm{b}=2200$, the moderately turbulent flow effects significant changes in the methanol production rate compared to the laminar flow. The mass flow rate recovered at the outlet is 56.4 \% higher in the turbulent case, implying a significant increase in methanol production due to turbulent transport. It is conceivable that this increase would be amplified by a longer duct than deployed here, and on the other hand, curtailed by a finite-rate chemistry, {\it{i.e.}} by a lower catalyst loading where the chemical time scales approach the (turbulent) flow time scales.

\begin{figure}[ht!]
\centering
 \includegraphics[width=0.98\textwidth]{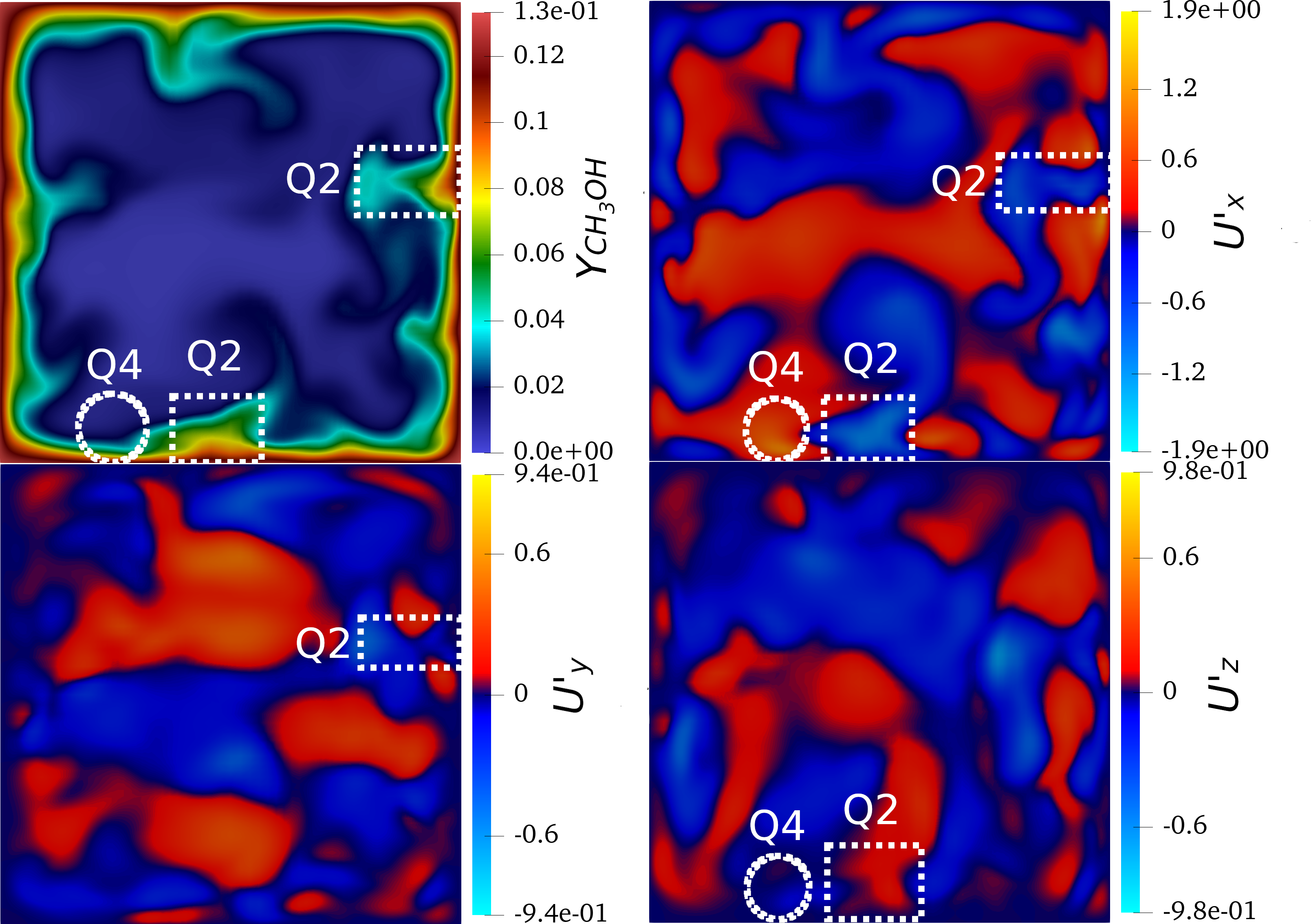}
 \caption{A: instantaneous methanol mass fraction at the outlet at $Re_\mathrm{b}=2200$. B, C and D: the streamwise, spanwise and transverse velocity fluctuations at the outlet at $Re_\mathrm{b}=2200$, respectively. Q2: site of an ejection event. Q4: site of a sweep event.}
 \label{fig:fig8}
\end{figure}

\begin{table}[t!]
    \centering
    \begin{tabularx}{0.95\textwidth}{>{\hsize=.75\hsize}X|>{\hsize=.75\hsize}X|>{\hsize=.75\hsize}X|X}
         Quadrant & Streamwise velocity fluctuation & Wall-normal velocity fluctuation & Explanation \\ \hline
         Q1 & $> 0$ & $> 0$ & Wall-to-center transport of high momentum fluid. \\
         Q2 - ejection & $< 0$ & $> 0$ & Wall-to-center transport of low momentum fluid. \\
         Q3 & $< 0$ & $< 0$ & Center-to-wall transport of low momentum fluid. \\
         Q4 - sweep & $> 0$ & $< 0$ & Center-to-wall transport of high momentum fluid. \\
    \end{tabularx}
    \caption{The 4 quadrant structures as defined by the sign of the streamwise and wall-normal velocity fluctuations. Physical descriptions of the structures are also provided.}
    \label{tab:tab4}
\end{table}
The net effect of turbulent transport on the local reaction maxima and minima rate is observed in ejection/sweep structures formed in pairs as defined in the context of a quadrant analysis~\citep{joung2007direct}. In wall-bounded flows, these structures can be identified based on the sign of the local fluctuating component of both the streamwise and wall-normal velocity components in the proximity of the wall. Therefore, four distinct combinations can be generated and these are summarized in Table~\ref{tab:tab4}, of which ejection (Q2) and sweep (Q4) structures are of interest here. Physically in the ejection region (Q2), corresponding to a negative streamwise velocity fluctuation and a fluctuation directed away from the wall in the wall-normal component, low momentum fluid is uplifted towards the center of duct witnessed as the bulging of the methanol boundary layer towards the center at these locations (Q2 in Fig.~\ref{fig:fig8}). In the sweep region (Q4), a positive streamwise fluctuation and a fluctuation directed towards the wall in the wall-normal component occurs, and the boundary layer is compressed instead (Q4 in Fig.~\ref{fig:fig8}).

\begin{figure}[ht!]
\centering
 \includegraphics[width=0.98\textwidth]{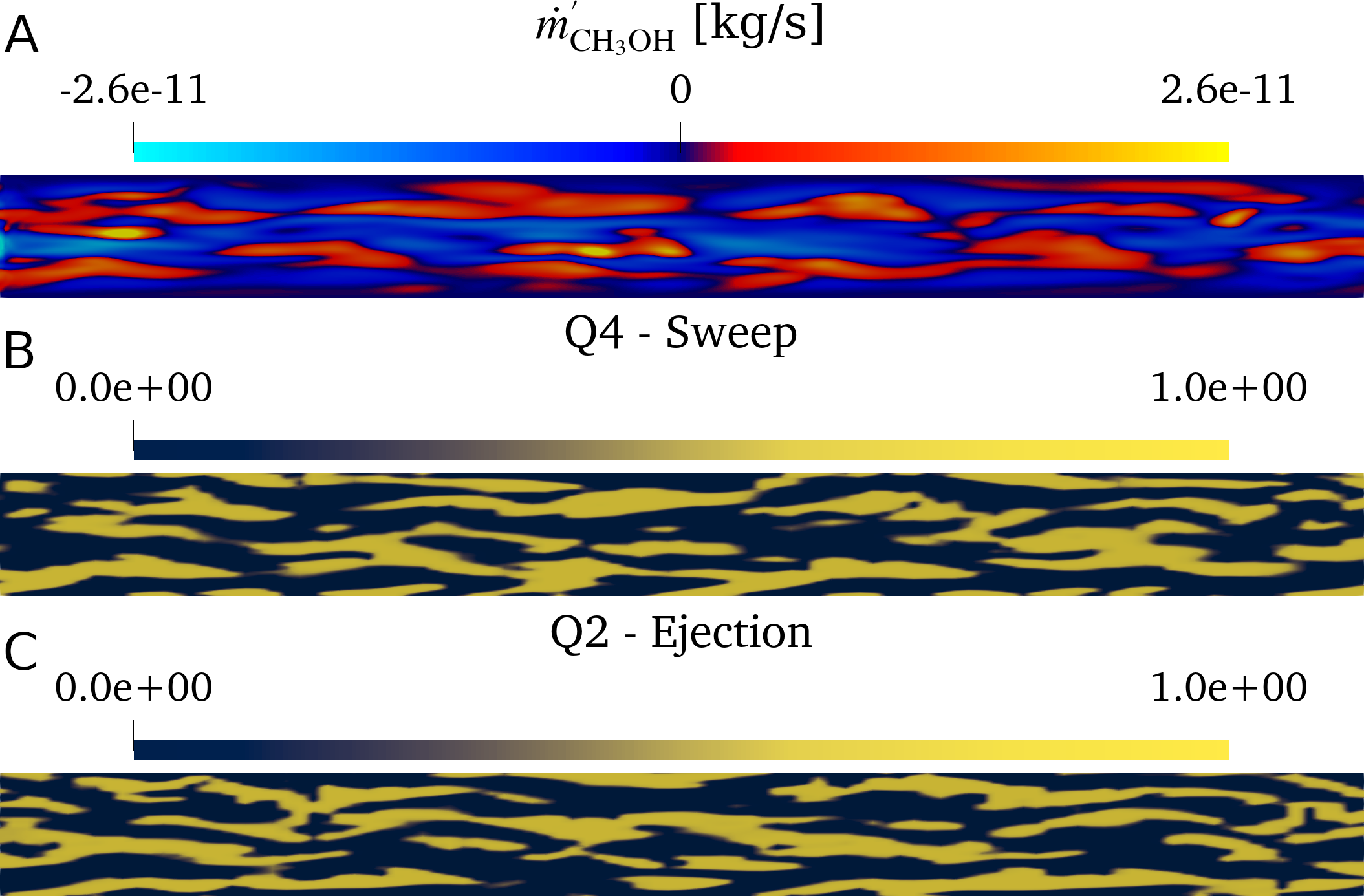}
 \caption{$Re_\mathrm{b}=2200$ - A: methanol formation rate fluctuations at a single catalytic wall ($xz$-plane, $y=-h$). The regions of B: sweep and C: ejection events demarcated by a binary colormap at a wall-normal distance of 10 wall-units ($y^+=10$) from the wall depicted in A. 1 (0) means an ejection/sweep structure is present (not present) based on the sign of the velocity fluctuations as explained in Table~\ref{tab:tab4}.}
 \label{fig:fig9}
\end{figure}
Fig.~\ref{fig:fig9} illustrates the correlation between the sweep (Fig.~\ref{fig:fig9} B) and ejection (Fig.~\ref{fig:fig9} C) structures with the local maxima and minima in the instantaneous methanol formation rate fluctuations $\dot{m}^{\prime}_{\mathrm{CH}_3\mathrm{OH}}$ (Fig.~\ref{fig:fig9} A). Note, that $\dot{m}_{\mathrm{CH}_3\mathrm{OH}}$ is evaluated by locally integrating $\dot{\omega}_{\mathrm{CH}_3\mathrm{OH}}$ over the cell volumes neighboring the catalytic wall and is thus distinct from mass flow rate at the outlet discussed earlier. The structures are evaluated at a distance of 10 wall-units ($y^+=10$, see the definition in ~\ref{appx:appx1}) from a single catalytic wall ($xz$-plane, $y=-h$) based on the velocity fluctuations as outlined in Table~\ref{tab:tab4}. This distance corresponds to the area of the highest turbulent activity in the setup and a value of 1 (0) indicates that the fluctuation condition for the sweep/ejection event is met (not met). The sweep zones (yellow) in Fig.~\ref{fig:fig9} B correlate reasonably well with the reaction hotspots (red, yellow) in A while similarly, the ejection zones (yellow) in C show similar patterns to those of the reaction minima (blue, cyan) in A. Physically, this implies that at the ejection sites, the ejected fluid is replaced by inflow from the neighboring sweep regions which in turn are replenished with fresh reactant from the duct center, producing local reaction maxima in methanol production. At the ejection site, however, the reactions are diminished due to insufficient reactant transport to the catalytic walls, resulting in local methanol formation rate minima.

\begin{figure}[ht!]
\centering
 \includegraphics[width=0.98\textwidth]{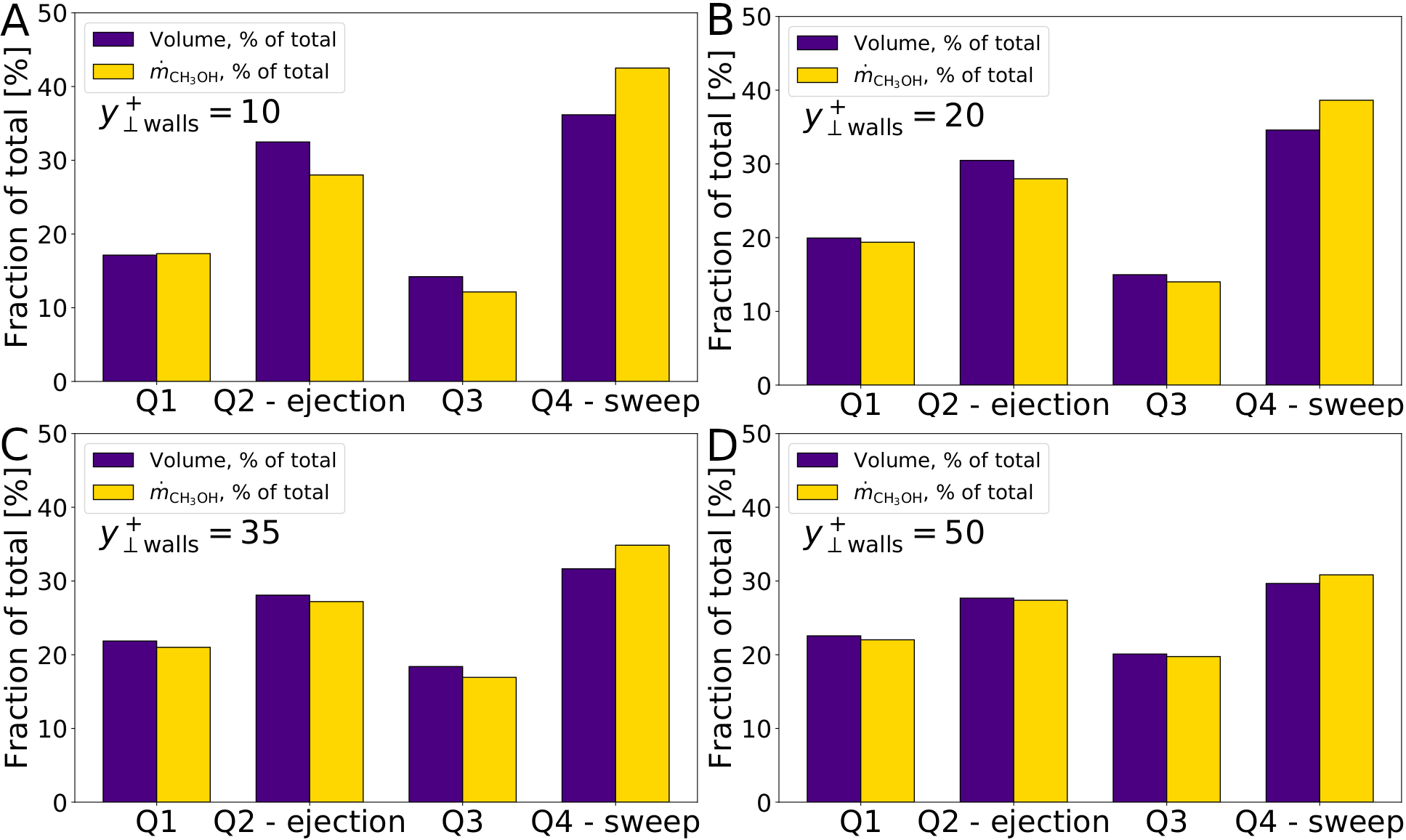}
 \caption{The fraction of total volume and methanol formation rate by each quadrant as defined at A: $y^+=10$, B: $y^+=20$ , C: $y^+=35$ or D: $y^+=50$ from the catalytic walls.}
 \label{fig:fig11}
\end{figure}

Further, the relative contribution of each quadrant to the total methanol formation rate in the reacting zone at $Re_\mathrm{b}=2200$ is presented in Fig.~\ref{fig:fig11}. Here, this quantity is obtained by summing $\dot{m}_{\mathrm{CH}_3\mathrm{OH}}$ over the cells neighboring the catalytic walls in the reacting zone. Over each catalytic wall, a plane is then overlayed to identify and match each local cell value $\dot{m}_{\mathrm{CH}_3\mathrm{OH}}$ at the catalytic wall with one of the four quadrants at the plane as demonstrated for sweep and ejection regions in Fig.~\ref{fig:fig9} B and C. These planes are defined either at $y^+=10$, 20, 35 or 50 wall units apart from the catalytic walls. The quadrant identification planes located at $y^+=10$, 20, 35 are beyond the viscous sublayer yet within the buffer layer, where the majority of the turbulent kinetic energy and structures are generated. However, the plane at $y^+=50$ is in the bulk turbulent flow region. This identification allows for binning the total $\dot{m}_{\mathrm{CH}_3\mathrm{OH}}$ according to the four quadrants and the relative fraction ($\dot{m}_{\mathrm{CH}_3\mathrm{OH}}$, \% of total) produced by each quadrant is presented in Fig.~\ref{fig:fig11} A, B, C and D for the identification planes located at $y^+=10$, 20, 35 or 50, respectively. At each plane, the total volume $V$ of the cells coinciding with that plane is also binned to each quadrant. The relative contribution of each quadrant to the total volume spanned by the planes then represents the relative incidence of each quadrant in the reacting zone. 

Judging by Figs.~\ref{fig:fig11} A, B, C, the sweeping regions clearly contribute the majority of the methanol produced in the reacting zone of the duct, while catalytic wall regions associated with overlaying Q3 structures contribute the least. Importantly, this result is robust and independent of the location of the plane on the condition that the plane is located in the buffer layer. However, with increasing wall-distance, the turbulent structures become decoupled from the chemistry transpiring at the catalytic walls, and this result becomes particularly difficult to perceive outside the buffer layer (Fig.~\ref{fig:fig11} D). Evidently, the mean volumetric methanol formation rate conditional on Q4, {\it{i.e.}} $\dot{m}_{\mathrm{CH}_3\mathrm{OH | Q4}}$ / $V_\mathrm{| Q4}$ in catalytic wall regions associated with Q4 (sweep) structures, is also larger than the unconditional mean volumetric methanol formation rate ($\sum_{i}$ $\dot{m}_{\mathrm{CH}_3\mathrm{OH | i}}$ / $V_\mathrm{| i}$, $i=$ Q1, Q2, Q3, Q4) in the reacting zone. This can be seen in Figs.~\ref{fig:fig11} A, B and C, where the relative formation rate is consistently higher than the relative volume occupied by Q4 (sweep) structures. Contrarily, the mean volumetric methanol formation rate conditional on Q2 (ejection) is lower than the unconditional mean. As a remark, these findings are consistent with the previous discussion of Fig.~\ref{fig:fig9}, where local maxima (minima) in $\dot{m}^{\prime}_{\mathrm{CH}_3\mathrm{OH}}$ where associated with sweep (ejection) structures.

\section{Conclusions}
In this study, the production of synthetic methanol (\ce{CH3OH}) in an isothermal duct coated with catalytic material was examined by full 3D CFD simulations. Two laminar and two turbulent cases were studied, corresponding to (bulk) Reynolds numbers $Re_\mathrm{b}=100$, 500, 1100 and 2200. The simulations were scale-resolved up to the DNS limit, and the kinetic model by Graaf~{\it{et al.}}~\citep{kiss2016novel} for methanol synthesis from \ce{CO2} in the presence of \ce{H2} on catalytic surfaces was employed. The thermodynamic variables were set to typical temperature ($T=503.15$ K) and pressure ($p = 5$ MPa) values utilized in commercial reactors while the reactant ratio $R=$ \ce{H2}:\ce{CO2} was set to $R=3$ (stoichiometric ratio). Additionally, a high catalyst loading was employed, yielding a space-velocity of 0.086, 0.44, 0.96 and 1.93 m$^3$/(kg$_\mathrm{cat}$ h) for the outlined Reynold numbers, respectively.

The single-pass conversion efficiency of \ce{CO2} and yield \ce{CH3OH} decreased with increasing $Re_\mathrm{b}$ due to decreased residence time of the reactants at the catalytic surfaces. However, the methanol mass flow rate reported at the outlet increased since this is offset by the increased mass flux of reactants at higher $Re_\mathrm{b}$. The employed kinetic model predicted negligible direct hydrogenation of \ce{CO2} in the present setup and methanol was hydrogenated from \ce{CO} originating from the reverse water-gas shift reaction. The steady-state Sherwood number and mass transfer coefficients for methanol were reported for each $Re_\mathrm{b}$, and the effect of turbulence on methanol yield was evaluated for the two turbulent cases. While no appreciable improvement in methanol production was observed in the marginally turbulent flow ($Re_\mathrm{b}=1100$), a significant increase of 56.4 \% in methanol mass flow rate at outlet was observed for the moderately turbulent flow ($Re_\mathrm{b}=2200$) as compared to a laminar flow with corresponding Reynolds number. This increase was due to improved reactant transport to the catalytic walls under the transport-limited chemistry. Furthermore, the local methanol formation rate exhibited minima/maxima at catalytic wall locations corresponding to pairs of ejection/sweep regions adjacent to the walls, respectively. Due to the negligible pressure drop in the duct geometry and lack of heat-transfer from the walls (isothermal setup), turbulence had no considerable impact on the chemical equilibrium and, thus, relative reaction rates.

These results provide quantitative insight into the effects of turbulence on synthetic methanol production and yield in square ducts under isothermal, fast chemistry conditions that can be potentially extrapolated to assess monolithic reactor performance in turbulent flow conditions. The results suggest that turbulent flow may be utilized to enhance methanol production and throughput in such setups, albeit at the cost of further recycling of the reactants. Further study incorporating a suitable turbulence model to the Reynolds-averaged Navier-Stokes equations (RANS) would allow for assessing the impact of pressure, temperature and reactant ratio on the performance characteristics of the reactor subjected to turbulent flow.

\section*{Acknowledgements}
\noindent We thank the Research Council of Finland (grant No. 335516) and Business Finland (grant 8797/31/2019) for their financial support. We also wish to extend our thanks to the Aalto Science-IT project for the high-performance computational resources. The authors also wish to acknowledge CSC – IT Center for Science, Finland, for the generous computational resources.

\bibliographystyle{elsarticle-num-names}
\bibliography{master}

\begin{appendix}
\section{Validation} \label{appx:appx1}
\begin{figure}[ht!]
\centering
 \includegraphics[width=0.98\textwidth]{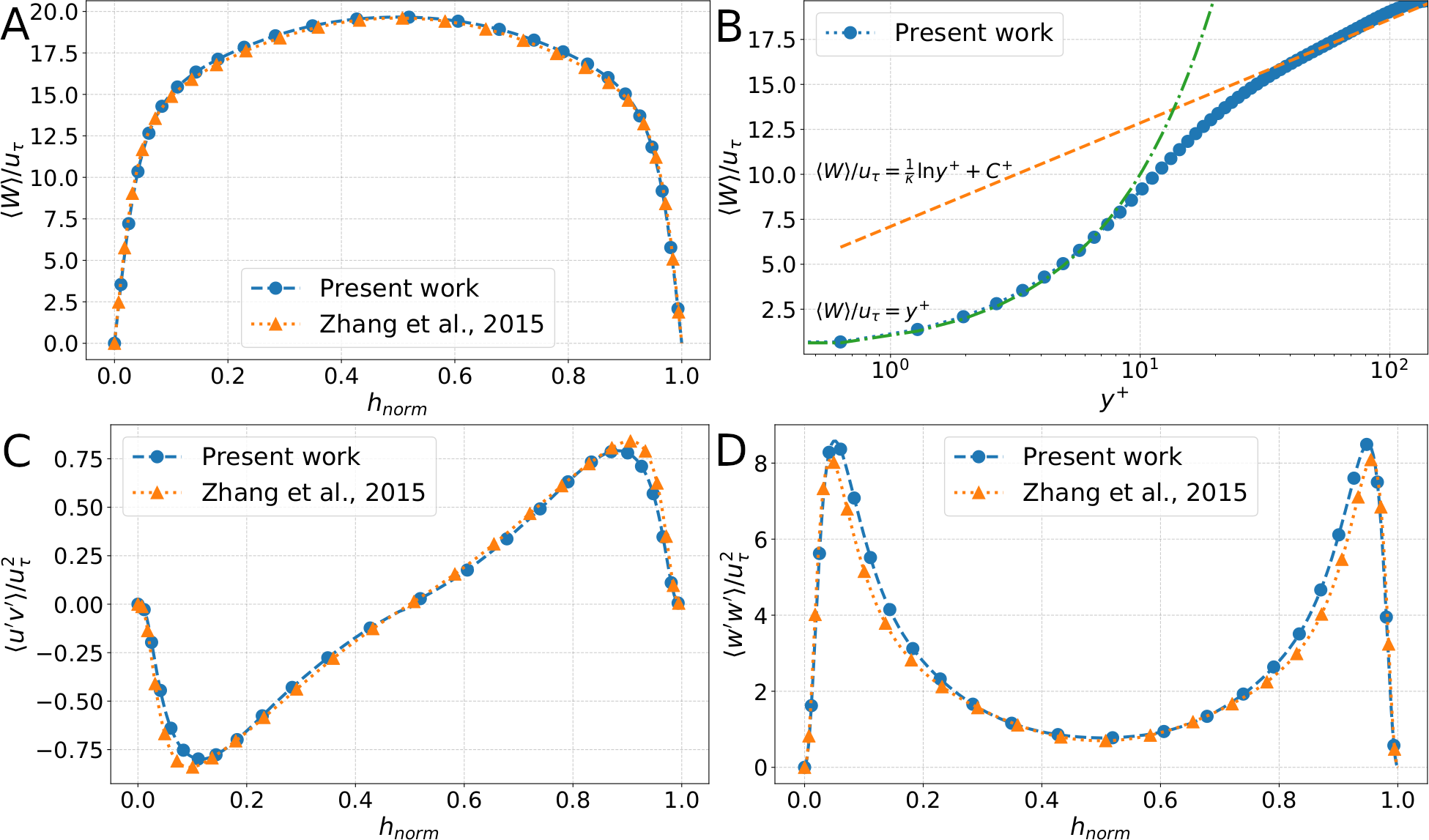}
 \caption{A comparison of the present DNS of a non-reacting, incompressible flow in a duct with the results of Zhang {\it{et al.}} at $Re_{\tau} = 150$. A: The time-averaged streamwise velocity at the duct centerline and B: the same profile plotted as a function of the wall coordinates $y^+$. C: The time-averaged Reynold stress components $\langle U'V'\rangle$ and D: $\langle W'W'\rangle$ at the duct centerline.}
 \label{fig:figappx1}
\end{figure}

\begin{table}[t!]
    \centering
    \begin{tabular}{l|c|r} 
         Property & Present simulations & Zhang {\it{et al.}} \\ \hline
         $Re_{\tau}$ & 150 & 150 \\
         $u_{\tau}$ & 0.22 & 0.22 \\
         $N_x \times N_y \times N_z $ & 128 $\times$ 128 $\times$ 256 & 128 $\times$ 128 $\times$ 160 \\
         $L / h$ & 12 & $2 \cdot \pi$ \\
         min $\Delta y^+$ & 0.48 & 0.224 \\
         max $\Delta x^+ / \Delta y^+$ & 11.00 & 52.589 \\
         CPUs & 256 & 32 \\
    \end{tabular}
    \caption{A comparison of the mesh properties between the present validation study and Zhang and colleagues~\citep{zhang2015direct}.}
    \label{tab:taba2}
\end{table}
Here, a validation of the numerical schemes and the mesh are presented by simulating an incompressible, non-reactive fluid flow in the recycling zone of the full 3D duct presented in Fig.~\ref{fig:fig1}. The DNS data of a duct flow provided by Zhang and co-workers~\citep{zhang2015direct} are employed as reference. Table~\ref{tab:taba2} provides details on the mesh parameters employed by both studies. Additionally, Fig.~\ref{fig:figappx1} presents a comparison between the results of the present simulation framework and the reference at $Re_{\tau}=150$. $Re_\tau = (h/2) u_\tau / \nu$ is the friction Reynolds number, where $h/2$ is the duct half-width, $u_\tau$ is the friction velocity at wall and $\nu$ is the kinematic viscosity. The friction velocity is defined as $u_\tau = \sqrt{\tau_w / \rho}$, where $\tau_w$ is the wall shear stress and $\rho$ is the density. In the present case, $Re_\tau = 150$ corresponds to a bulk Reynolds number of $Re_{\mathrm{b}}=2200$, which is also the highest $Re_\mathrm{b}$ value employed in the reactive flow simulations in this study.

The time-averaged (streamwise) velocity defined at the duct central plane and normalized with $u_\tau$ is first illustrated in Fig.~\ref{fig:figappx1} A, indicating an excellent agreement. Furthermore, the slip velocity at the wall reported in~\citep{zhang2015direct} ($u_\tau = 0.22$) is also captured with the present simulation with an identical value. Fig.~\ref{fig:figappx1} B depicts the normalized mean velocity at half-channel width as a function of the $y^+$ values, where $y^+ = y u_\tau / \nu$ are the normalized wall units. Importantly, 1) the first near-wall cell has an $y^+$ value of 0.5, 2) the successive cells are stretched with a constant factor of approximately 1.04 and 3) the total amount of grid points in the viscous sublayer number 18. These imply that the mesh is fine enough for resolving the turbulent length scales at DNS-level at this Reynolds number and below. Fig.~\ref{fig:figappx1} C displays the transverse-spanwise component of the (mean) Reynold stress normalized by $u_\tau^2$ and plotted over the normalized duct width $h_{\mathrm{norm}}$. Here, the values correspond well to each other in centre of the duct though the extrema in the buffer layer slightly differ. In Fig.~\ref{fig:figappx1} D, similar data is depicted for the streamwise-streamwise component of the Reynold stress indicating a good agreement well with the reference data although slight deviations are again present in the maxima. The minor differences are likely the result of the higher order discretization schemes employed in~\citep{zhang2015direct} and discrepancies in the maximum aspect ratio of the mesh as the maximum aspect ratio in this study is 11.0 whereas in~\citep{zhang2015direct} it is reported as 52.59.

\end{appendix}

\end{document}